\documentclass[12pt]{article}
\usepackage{graphicx}

\begin{document}
\begin{center}
\textbf{Quantum Fluctuations of the Gravitational Field and Propagation of
Light: a Heuristic Approach\footnote{Talk presented at
Qed 2000, 2nd Workshop on Frontier Tests of Quantum Electrodynamics
and Physics of the Vacuum.}.}
\end{center}

\begin{center}
Stefano Ansoldi$^{*}$ and \underline {Edoardo Milotti}$^{**}$
\end{center}

\begin{center}
${}^{(*)}$\textit{Dipartimento di Fisica Teorica dell'Universit\`a di Trieste\\
and I.N.F.N. - Sezione di Trieste,}\\
\textit{Strada Costiera, 11 - I-34014 Miranare - Trieste, Italy}\\
{\tt{}e-mail: ansoldi@trieste.infn.it}
\end{center}

\begin{center}
${}^{(**)}$\textit{Dipartimento di Fisica dell'Universit\`a di Udine\\
and I.N.F.N. - Sezione di Trieste,}\\
\textit{Via delle Scienze, 208 - I-33100 Udine, Italy}\\
{\tt{}e-mail: milotti@fisica.uniud.it}
\end{center}

\begin{abstract}
\noindent
Quantum Gravity is quite elusive at the experimental
level; thus a lot of interest has been raised by recent searches for quantum
gravity effects in the propagation of light from distant sources, like gamma
ray bursters and active galactic nuclei, and also in earth-based
interferometers, like those used for gravitational wave detection. Here we
describe a simple heuristic picture of the quantum fluctuations of the
gravitational field that we have proposed recently, and show how to use it
to estimate quantum gravity effects in interferometers.
\end{abstract}

\section{Introduction}
\label{sec:1}

\begin{figure}
\begin{center}
%*[bbllx=0.15in,bblly=1.31in,bburx=10.19in,bbury=11.34in,scale=0.86]
\includegraphics[scale=0.7]{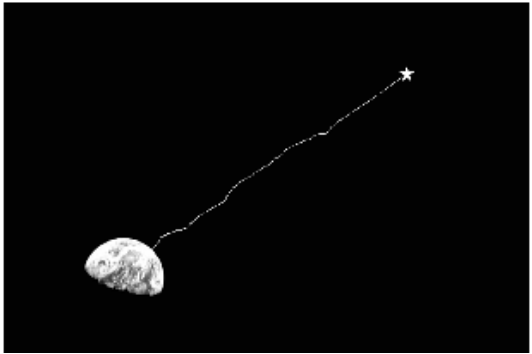}
\caption{{\small{}In addition to atmospheric scintillation we expect
scintillation effects due to the background of gravitational waves that
affect starlight as it travels to the earth. The light emitted by a star
does not travel in a straight line, but is scattered by this background:
therefore it displays angular displacement and the time of arrival
fluctuates randomly.}}
\label{fig:1}
\end{center}
\rule{5.5cm}{0.5pt}
\end{figure}

The propagation of light in random or fluctuating media has long been used
as a probe of their statistical properties \cite{bib:1}. This holds true also for
exotic media like the background of gravitational waves in the space between
the earth and some faraway light source \cite{bib:2,bib:3,bib:4}
(see figure \ref{fig:1}).

Recently several authors have proposed searches for quantum gravity effects
in the propagation of light over cosmological distances or in earth-based
interferometers; for instance, according to Ellis, Mavromatos and Nanopoulos
\cite{bib:5}, the vacuum of quantum gravity should be dispersive and this should show
up as an energy-dependent spread in the arrival times of energetic photons
from distant sources like gamma-ray bursters, active galactic nuclei, and
gamma-ray pulsars. This opens up interesting observational opportunities for
experimentalists (see, e.g. \cite{bib:6,bib:7}), but requires either a violation of
Lorentz invariance or of the equivalence principle \cite{bib:8}.

The proposal \cite{bib:5} is based on the latest developments of $D$-brane theory, and
it is yet another picture of the ``space-time foam'' first proposed by Wheeler
in 1957 \cite{bib:9}, and later considered by other authors such as Hawking \cite{bib:10}.

Here we propose a heuristic treatment of these background fluctuations,
which is based on the direct use of the uncertainty principle for energy and
time, with additional assumptions on the dynamic and statistical behaviour
of the resulting quantum fluctuations. Afterwards we use this model to
evaluate the spectral properties of the fluctuations of the energy density
and of the gravitational potential, and to derive the behaviour of light in
a generic two-arm interferometer.

P. Bergmann \cite{bib:2} originally suggested to search for low-frequency components
of gravitational waves by attempting to detect fluctuations in the intensity
of light, but actually most searches for this effect have been carried out
looking for fluctuations in the arrival times of pulses from millisecond
pulsars and other such sources. See \cite{bib:3} for the details of how one such
search is performed, and \cite{bib:4} for a recent review.

\section{Fluctuations of the\\ gravitational energy density}
\label{sec:2}

We consider now a single quantum fluctuation of the vacuum energy density
and assume that it is uniformly spread over a spherical region, then the
total energy associated with the fluctuation is
\begin{equation}
\label{eq1}
E = \frac{{4\pi R_{0}^{3}} }{{3}}\rho ,
\end{equation}
where $R_{0} $ is the radius of the bubble. We assume that the bubble
expands at the speed of light, so that $R_{0} = ct$, where $t$ is the
time of creation of the bubble, and that it satisfies the time-energy
uncertainty principle at all times\footnote{Quantum mechanics enters this
simple model only by way of the uncertainty principle, but it does so
``peacefully'', and coexists with relativistic invariance in a rather
natural way.}, so that $E \approx \hbar t$, and therefore we find
\begin{equation}
\label{eq2}
\rho \approx \frac{{3\hbar} }{{4\pi c^{3}t^{4}}}.
\end{equation}
It is also important to note that the expanding bubbles are just a
representation of the light cones in $3$-space, therefore they are Lorentz
invariant, i.e., they would look the same in any other Lorentz-boosted
reference frame. A rough form of Lorentz invariance is thus present in our
model, whose causal structure is compatible with a relativistic model of
spacetime, and these fluctuations satisfy the requirement, first discussed
by Zeldovich \cite{bib:12}, that the quantum vacuum must indeed be Lorentz invariant.

\begin{figure}
\begin{center}
%*[bbllx=0.15in,bblly=1.31in,bburx=3.22in,bbury=4.38in,scale=0.80]
\includegraphics[scale=0.6]{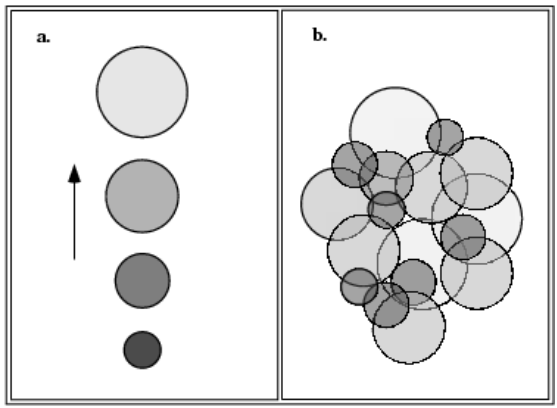}
\caption{\small{\texttt{a.} We picture the fluctuations of the gravitational
energy density as ``bubbles'' that expand at the speed of light, and are
uniformly filled with a decreasing energy density that satisfies at all
times the time-energy uncertainty principle; \texttt{b.} an observer sees the whole
of space seething with bubbles, whose number density in a given space-time
volume follows a simple Poisson statistics. More pictures and simulations of
these fluctuations can be found on the web \cite{bib:11}.}}
\label{fig:2}
\end{center}
\rule{5.5cm}{0.5pt}
\end{figure}

Now let $n\left( {\vec{x},t} \right)$ be the number density of the fluctuations
that occur in spacetime, i.e., $n\left( {\vec{x},t} \right)dVdt$
fluctuations are
created in a small space-time volume $dVdt$ at position $\vec{x}$ and time
$t$, and assume that $n\left( {\vec{x},t} \right)dVdt$ is a Poisson variate
with average $n_{0} dVdt$ and variance $n_{0} dVdt$ (see fig. \ref{fig:2});
then the
average density observed at $\left( {\vec{x}_{0} ,t} \right)$ and due to the
fluctuations created at a distance \textit{r} and at an earlier time
$t'$ is\footnote{Here we assume that the energy densities are
sufficiently small so that they can indeed be added linearly.}
\begin{equation}
\label{eq3}
d^{2}\rho = \frac{{3\hbar n_{0}} }{{c^{3}\left( {t - t_{0}}
\right)}}r^{2}drdt.
\end{equation}
Then the total energy density observed at $\left( {\vec{x}_{0} ,t} \right)$ and
due to all the prior fluctuations in the light-cone of the observer is given
by
\begin{equation}
\label{eq4}
\rho _{\mathrm{tot}} \left( {t} \right) \approx \int_{r_{\mathrm{min}}} ^{r_{\mathrm{max}}}  {dr\int_{
- T_{0}} ^{t - r/c} {dt_{0}\frac{{3\hbar n_{0}} }{{c^{3}\left( {t - t_{0}}
\right)}}r^{2}}}  \approx \hbar n_{0}
\ln\frac{{r_{\mathrm{max}}} }{{r_{\mathrm{min}}} },
\end{equation}
where $T_{0}$ is the age of the Universe, and $r_{\mathrm{min}} $, $r_{\mathrm{max}} $ are the
minimum and the maximum distance from the observer. Furthermore in (\ref{eq4}) we
have dropped the $\vec{x}$ dependence, because we assume translational
invariance. The maximum distance can be taken to be the present radius of
the Universe $r_{\mathrm{max}} \approx {{c} \mathord{\left/ {\vphantom {{c} {H}}}
\right. \kern-\nulldelimiterspace} {H}}$, while the minimum radius
corresponds to the smallest bubble that can possibly be observed, that is
that bubble that emerges from a ``mini black hole'' stage and turns into a
``normal'' fluctuation, so that $r_{\mathrm{min}} $ is just the Schwartzschild radius
of the fluctuation:
\begin{equation}
\label{eq5}
r_{\mathrm{min}} = \frac{{2G}}{{c^{2}}} \cdot \frac{{E}}{{c^{2}}} \approx
\frac{{2G}}{{c^{2}}} \cdot \frac{{\hbar} }{{c^{2}t}} = \frac{{2G\hbar
}}{{c^{3}r_{\mathrm{min}}} }.
\end{equation}
We solve eq. (\ref{eq5}) and find
\begin{equation}
\label{eq6}
r_{\mathrm{min}} = \sqrt {\frac{{2G\hbar} }{{c^{3}}}} ,
\end{equation}
which is just the Planck length. Thus eq. (\ref{eq4}) gives an average density
\begin{equation}
\label{eq7}
\rho _{\mathrm{tot}} \approx \frac{{1}}{{2}}\hbar n_{0} \ln\frac{{c^{5}}}{{2G\hbar
H^{2}}} \quad .
\end{equation}
It is worthwhile to notice that a minimum and a maximum scale show up
naturally in the calculations, and that the total energy density (\ref{eq7}) is
seamlessly related to both the micro and the macrostructure of the Universe.

The same formalism can be used to estimate the variance of the energy
density fluctuations, and proceeding as before one finds
\begin{equation}
\label{eq8}
\sigma _{\rho} ^{2} = \frac{{9\hbar ^{2}cn_{0}} }{{28\pi} }\int_{r_{\mathrm{min}}
}^{r_{\mathrm{max}}}  {\frac{{dr}}{{r^{5}}}} = \frac{{9\hbar ^{2}cn_{0}} }{{112\pi
}}\left( {\frac{{1}}{{r_{\mathrm{min}}^{4}} } - \frac{{1}}{{r_{\mathrm{max}}^{4}} }} \right)
\approx \frac{{9c^{7}n_{0}} }{{448G^{2}\pi} }
.
\end{equation}
It is also important to notice that the average energy density (\ref{eq7})
contributes to the effective cosmological constant \cite{bib:13}:
\begin{equation}
\label{eq9}
\Delta \Lambda \left( {t} \right) = \frac{{8\pi} }{{c^{4}}}G\rho _{\mathrm{tot}}
\approx \frac{{8\pi} }{{c^{4}}}G\hbar n_{0} \ln\frac{{c}}{{\ell _{P} H\left(
{t} \right)}};
\end{equation}
recent observations favour a nonzero and positive cosmological constant
\cite{bib:14,bib:15} such that $\Omega _{\Lambda}  \approx 0.7$ and $\Lambda = 3\Omega
_{\Lambda}  {{H^{2}} \mathord{\left/ {\vphantom {{H^{2}} {c^{2}}}} \right.
\kern-\nulldelimiterspace} {c^{2}}}$; this means that we can use (\ref{eq9}) to
estimate the order of magnitude of the free parameter $n_{0} $, if we assume
that most of the effective cosmological constant is due to the fluctuations
considered here:
\begin{equation}
\label{eq10}
n_{0} \approx \frac{{3\Omega _{\Lambda}  H^{2}c^{2}}}{{4\pi G\hbar \ln\left(
{{{c^{5}} \mathord{\left/ {\vphantom {{c^{5}} {2G\hbar H^{2}}}} \right.
\kern-\nulldelimiterspace} {2G\hbar H^{2}}}} \right)}} \approx 3 \cdot
10^{22}\,\mathrm{m}^{ - 3}\mathrm{s}^{ - 1}.
\end{equation}

\section{Fluctuations of the gravitational potential}
\label{sec:3}

Since we assume that fluctuations spread with the speed of light, there can
be no gravitational potential outside the bubble, because for an external
observer there is no way to know that it exists, while inside the bubble we
assume that the gravitational potential is just the (Newtonian) potential of
a sphere of uniform energy density $\rho $:
\begin{equation}
\label{eq11}
\Phi \left( {r} \right)
=
\cases{
    \displaystyle
	- \frac{4 \pi G \rho}{3 c^{2}}
	  \left(
	  		 \frac{3}{2} R _{0} ^{2}
	  		 -
			 \frac{1}{2} r^{2}
	  \right)
	\qquad
	&
	if $\quad r < R_{0}$
	\cr \cr
 	0
	&
	if  $\quad r > R_{0}$
	  }
	.
\end{equation}
Integrating as in the case of the energy density one finds:
\begin{equation}
\label{eq12}
\Phi _{\mathrm{tot}} \approx - \frac{{16\pi G\hbar n_{0}} }{{3c^{2}}}r_{\mathrm{max}}^{2}
\approx - \frac{{16\pi G\hbar n_{0}} }{{3cH}}
\end{equation}
and
\begin{equation}
\label{eq13}
\sigma _{\Phi} ^{2} = \frac{{68\pi G^{2}\hbar ^{2}n_{0}
}}{{35c^{3}}}\ln\frac{{r_{\mathrm{max}}} }{{r_{\mathrm{min}}} }
.
\end{equation}

\section{Correlation functions}
\label{sec:4}

A somewhat more complicated integration yields the correlation function of
the gravitational potential:
\begin{eqnarray}
&&
\left\langle {\Delta \Phi \left( {\vec{x}_{1} ,t_{1}}  \right)\Delta \Phi \left(
{\vec{x}_{2} ,t_{2}}  \right)} \right\rangle = \left( {\frac{{G\hbar
c}}{{2c^{2}}}} \right)^{2}n_{0} \int\limits_{S} d^{3}x\int\limits_{t < t ^{\star}
} dt\left( {3 -
\frac{{r_{1}^{2}}}{{c^{2}\left( {t_{1} - t} \right)^{2}}}} \right) \times
\nonumber \\
&&\qquad
 \times \left(
\frac{1}{c^{2}\left( t_{1} - t \right)^{2}} \right)
\left( {3 - \frac{{r_{2}^{2}} }{{c^{2}\left( {t_{2} - t}
\right)^{2}}}} \right)\left( {\frac{{1}}{{c^{2}\left( {t_{2} - t}
\right)^{2}}}} \right) ,
\label{eq14}
\end{eqnarray}
where the shorthand notation $r_{1,2} = \left| {\vec{x}_{1,2} - \vec{x}} \right|$ has
been used,
$$t ^{\star} = \min \left( t_{1} - r_{1}/c , t_{2} - r_{2} /c \right)$$
and \textit{S} is the range of acceptable values of $\vec{x}_{1,2} $ ---
i.e., values such that $r_{\mathrm{min}} < r_{1,2} < r_{\mathrm{max}} $.

The expression (\ref{eq14}) can be calculated for different special cases, e.g., the
time correlation function at a given space point is
\begin{equation}
\label{eq15}
R\left( {\tau}  \right) = \left\langle {\Delta \Phi \left( {\vec{x},0}
\right)\Delta \Phi \left( {\vec{x},\tau}  \right)} \right\rangle \approx
\frac{{68}}{{35}}\frac{{G^{2}\hbar ^{2}\pi} }{{c^{3}}}n_{0} \ln\frac{{r_{\mathrm{max}}
+ c\tau} }{{r_{\mathrm{min}} + c\tau} }
\end{equation}
and then, using the Wiener-Kintchine theorem, one finds
\begin{equation}
\label{eq16}
\left\langle {\left| {\Delta \Phi \left( {f} \right)} \right|^{2}}
\right\rangle = \int_{ - \infty} ^{ - \infty}  {R\left( {\tau}  \right)
\mathrm{e}^{ -
2\pi \mathrm{i} f\tau} d\tau}  \approx \frac{{34}}{{35}}\frac{{G^{2}\hbar ^{2}n_{0}
\pi} }{{c^{3}}} \cdot \frac{{1}}{{f}}
.
\end{equation}

\section{Irradiance fluctuations \\ in a two-arm interferometer}
\label{sec:5}

In the weak field approximation the frequency of light is linearly related
to the gravitational potential
\begin{equation}
\label{eq17}
\nu \left( {\vec{x},t} \right) \approx \nu _{0} \left( {1 - \frac{{\Phi \left(
{\vec{x},t} \right)}}{{c^{2}}}} \right)
;
\end{equation}
therefore, using the correlation function (\ref{eq14}), one finds that the time
correlation function for the irradiance fluctuations in a two-arm
interferometer is
\begin{eqnarray}
R_{I} \left( {\tau}  \right) & = & \left\langle {\Delta I\left( {t}
\right)\Delta I\left( {t + \tau}  \right)} \right\rangle
\nonumber \\
& = & \frac{{4I_{1} I_{2} \nu _{0}^{2} sin^{2}\varphi _{0}
}}{{c^{6}}}\int\!\!\!\int\limits_{L_{1} ,L_{2}}  d\ell _{1} d\ell _{2}
\times
\nonumber \\
& & \qquad \times
\left\langle {\Phi \left[ {\vec{x}\left( {\frac{{\ell _{1}} }{{c}}}
\right),\frac{{\ell _{1}} }{{c}}} \right]\Phi \left[ {\vec{x}\left( {\frac{{\ell
_{2}} }{{c}} + \tau}  \right),\frac{{\ell _{2}} }{{c}} + \tau}  \right]}
\right\rangle
,
\label{eq18}
\end{eqnarray}
where $\ell _{1,2} $ is the position along the path $L_{1,2} $, $\nu _{0} $
is the frequency of light in a zero-potential region, $I_{1,2} $ is the
irradiance of light along each path, and $\varphi _{0} $ is the average
phase difference.

We plan to complete the calculation of the time correlation function (\ref{eq18})
for specific interferometer designs in the near future.

\end{document}